\documentclass[conf]{new-aiaa}

\usepackage[utf8]{inputenc}
\usepackage{textcomp}
\usepackage{pgfplots}
\pgfplotsset{compat=newest}

\usetikzlibrary{positioning, fit, arrows.meta, shapes}
\usepackage{graphicx}
\usepackage{amsmath}
\usepackage{mathtools}
\usepackage[english]{babel}
\usepackage[version=4]{mhchem}
 \usepackage{amsmath}
\usepackage{siunitx}
\usepackage{longtable,tabularx}
 \usepackage{amsthm}
\usepackage[utf8]{inputenc}
\usepackage{graphicx}
\usepackage{subcaption}
\usepackage{amsmath}
\usepackage[version=4]{mhchem}
\usepackage{siunitx}
\usepackage{longtable,tabularx}
\usepackage{algorithm}
\usepackage{algpseudocode}
\usepackage{amsmath}
\usepackage{float}
\usepackage{titlesec}

\usepackage{xcolor}
\usepackage{textcomp}
\usepackage{graphicx}
\usepackage{amsmath}
\usepackage[version=4]{mhchem}
\usepackage{siunitx}
\usepackage{longtable,tabularx}
\usepackage{graphicx}
\usepackage{subcaption}
\usepackage{amsmath}
\usepackage{mathtools}
\usepackage[version=4]{mhchem}
\usepackage{siunitx}
\usepackage{longtable,tabularx}
\usepackage{algorithm}
\usepackage{algpseudocode}
\usepackage{amsmath}
\usepackage{amsfonts}
\usepackage{comment}
\usepackage{titlesec}
\usepackage{booktabs,subcaption,amsfonts,dcolumn}
\usetikzlibrary{backgrounds}
\usepackage{authblk}
\usepackage{amsmath} 
\usepackage{listofitems} 
\usetikzlibrary{arrows.meta} 
\usepackage[outline]{contour} 
\usepackage{xcolor}

\newcommand{\bu}{{\boldsymbol{u}}}

\newcommand{\bd}{{\boldsymbol{d}}}
\newcommand{\bomega}{{\boldsymbol{\omega}}}

\usepackage{dirtytalk}

\contourlength{1.4pt}

\usepackage{caption}
\usepackage{subcaption}

\colorlet{myred}{red!80!black}
\colorlet{myblue}{blue!80!black}
\colorlet{mygreen}{green!60!black}
\colorlet{myorange}{orange!70!red!60!black}
\colorlet{mydarkred}{red!30!black}
\colorlet{mydarkblue}{blue!40!black}
\colorlet{mydarkgreen}{green!30!black}
\colorlet{mypurple}{purple!80!black}
\colorlet{black}{black}

\usepackage{tikz}
\usetikzlibrary{positioning,arrows.meta,calc}
\usetikzlibrary{positioning, fit, calc, arrows.meta, patterns}
\usetikzlibrary{positioning, fit, calc, arrows.meta, patterns, shapes.misc}

\tikzset{
  >=latex, 
  node/.style={thick,circle,draw=myblue,minimum size=15,inner sep=0.5,outer sep=0.6},
  node in/.style={node,blue!20!black,draw=myblue!30!black,fill=myblue!25},
  node hidden/.style={node,purple!20!black,draw=mypurple!30!black,fill=mypurple!20},
  node convol/.style={node,orange!20!black,draw=mygreen!30!black,fill=mygreen!20},
  node out/.style={node,red!20!black,draw=mygreen!30!black,fill=mygreen!20},
  connect/.style={thick,black}, 
  connect arrow/.style={-{Latex[length=4,width=1]},thick,black,shorten <=0.5,shorten >=1},
  node 1/.style={node in}, 
  node 2/.style={node hidden},
  node 3/.style={node out}
}

\tikzset{
  arro/.style={
    ->,
    >=latex,
    line width=0.5mm 
  },
  bloque/.style={
    draw,
    minimum height=0.5cm,
    minimum width=0.5cm,
    line width=0.5mm 
  }  
}

\def\nstyle{int(\lay<\Nnodlen?min(2,\lay):3)} 

\title{Leveraging Gated Recurrent Units for Iterative Online Precise Attitude Control for Geodetic Missions}

\author[1]{Vrushabh Zinage\footnote{PhD Candidate, University of Texas at Austin, Austin, TX, USA, Student Member AIAA}}
\author[2]{Shrenik Zinage\footnote{Graduate Student, School of Mechanical Engineering, Purdue University, West Lafayette}}
\author[1]{Srinivas Bettadpur\footnote{Professor, Center for Space Research and Department of Aerospace Engineering and Engineering Mechanics,  University of Texas at Austin, Austin, TX, USA}}
\author[1]{Efstathios Bakolas\footnote{Associate Professor, Department of Aerospace Engineering and Engineering Mechanics, Senior Member AIAA.}}

\affil[1]{The University of Texas at Austin, Austin, Texas, 78712}
\affil[2]{Purdue University, West Lafayette, Indiana, 47906}

\begin{document}

\maketitle

\begin{abstract}
In this paper, we consider the problem of precise attitude control for geodetic missions, such as the GRACE Follow-on (GRACE-FO) mission. Traditional and well-established control methods, such as Proportional-Integral-Derivative (PID) controllers, have been the standard in attitude control for most space missions, including the GRACE-FO mission. Instead of significantly modifying (or replacing) the original PID controllers that are being used for these missions, we introduce an iterative modification to the PID controller that ensures improved attitude control precision (i.e., reduction in attitude error). The proposed modification leverages Gated Recurrent Units (GRU) to learn and predict external disturbance trends derived from incoming attitude measurements from the GRACE satellites. Our analysis has revealed a distinct trend in the external disturbance time-series data, suggesting the potential utility of GRU's to predict future disturbances acting on the system. The learned GRU model compensates for these disturbances within the standard PID control loop in real time via an additive correction term which is updated at regular time intervals. The simulation results verify the significant reduction in attitude error, verifying the efficacy of our proposed approach.
\end{abstract}

\section{Introduction\label{sec:introduction}}

The GRACE Follow-On mission is a joint scientific project between the United States and Germany, continuing the approach initiated by the earlier GRACE mission \cite{tapley2004gravity_grace_1,kornfeld2019grace_grace_2}. The missions were implemented for NASA by the Caltech/Jet Propulsion Laboratory in cooperation with the German Research Centre for Geosciences (GFZ), with mission operations managed by the German Space Operations Centre (DLR-GSOC). The primary aim of this mission is to gather data to produce both constant and time-varying maps of the Earth's gravity field. In May 2018, the GRACE Follow-on mission successfully launched two satellites into a polar orbit approximately $491$ km above the Earth. This mission continues to measure the Earth's gravity field, focusing on how it changes over time, and provides radio occultation measurements as well. The two satellites maintained a distance between each other of $170$ to $270$ km, working as detectors in the Earth's gravity field. They use a microwave tracking system to measure their distance apart very precisely, up to $1\mu $m. Furthermore, a laser ranging interferometer \cite{landerer2020extending_grace_laser_1,abich2019orbit_grace_laser_2}, is included to test new technology. To achieve the best scientific results, the satellites need to point at each other very steadily and accurately while minimizing the effect of any disturbances. This requires highly precise attitude control of their orientations with respect to each other.

One of the critical factors that can affect the accuracy of measurements and consequently the derived data product is the spacecraft attitude error. Current state-of-the-art methods to control and correct this error primarily involve the use of Proportional-Integral-Derivative (PID) controllers. Although PID controllers have been effective and theoretically well established, future missions may have higher precision requirements than what PID controllers can achieve at present. Specifically, geodetic missions, such as GRACE or satellite gravity gradiometers, have stringent requirements for attitude control due to several critical factors discussed in \cite{rosen2021analysis_mitch_thesis}. Furthermore, these missions require precise information about the attitude for accurate scientific analysis, which can be achieved by minimizing the effects of angular rates on the inertial sensors and external disturbances acting on these satellites. To achieve this goal, one must be able to take into account these disturbances and actively compensate them in the designed controllers.

Traditional and well established disturbance observers such as Active Disturbance Rejection Control (ADRC) \cite{han2009pid_adrc,zhang2019design_adrc_2} that can estimate the external disturbance online are restricted to systems that have a certain structure in their governing dynamics. Furthermore, they are not able to fully exploit the observed trend in the disturbances if there are any. Toward this goal, neural networks have recently been used successfully for modeling engineered systems for a considerable period \cite{narendra1992neural,gonzalez1998identification,chen1990non,milano2002neural,rico1993continuous}. Among the various forms of neural networks, recurrent neural networks (RNNs) stand out for their ability to incorporate feedback loops, allowing the output of one neuron (as shown in Fig. \ref{fig:neuron}) to influence its subsequent input, which mirrors temporal dynamics. This characteristic makes RNNs particularly suitable for modeling sequential processes and has led to their frequent application in the forecasting of time series \cite{vlachas2018data, pan2018long, pathak2018hybrid, lu2018attractor}. A common challenge with traditional RNNs is their struggle with short-term memory, often attributed to the problem of vanishing gradients \cite{pascanu2013difficulty}. The problem of vanishing gradients arises when the gradient of the loss function approaches zero, making it almost impossible to update the weights of the network. This prevents the network from learning effectively from the training data. An effective solution to this problem is the use of long-short-term memory (LSTM) networks \cite{hochreiter1997long}, a specialized form of RNN designed to capture long-term dependencies. LSTMs are distinguished by their use of three gates, which, while improving their capability, also result in greater memory requirements due to an increased number of training parameters. An alternative that addresses this issue is the gated recurrent unit (GRU) \cite{Cho2014}, which simplifies the structure by using only two gates, thus offering faster computations. Numerous studies have been conducted to evaluate the predictive performance of these RNN variants and their effectiveness in the modeling of complex engineered systems \cite{zinage2022investigation, shopov2019identification}. For example, \cite{shopov2019identification} conducted a comparative analysis of different RNN architectures, including traditional RNNs, LSTMs and GRUs, in the context of predicting time series data. They showed superior performance of LSTM and GRU networks over conventional RNNs, with the GRU model, in particular, demonstrating significant advantages over both traditional RNN and LSTM models.

The main contributions of this paper are as follows. First, using the GRACE data product that consists of the time series data of the attitude error, we estimate the external additive disturbances acting on the spacecraft. Second, leveraging the observed trend in the disturbance time series data, we design a GRU based model to design a minimally modified PID controller that actively compensates for the external disturbances. 

The outline of the paper is as follows. Section \ref{sec:preliminaries_and_proposed_approach} delves into the preliminaries followed by the proposed approach. Section \ref{sec:results} discusses the numerical simulations, and the concluding remarks are succinctly presented in Section \ref{sec:conclusion}.
\section{Preliminaries and Proposed Approach\label{sec:preliminaries_and_proposed_approach}}
\subsection{Coordinate system}
\begin{figure}
     \centering
     \begin{subfigure}[b]{0.3\textwidth}
         \centering
         \includegraphics[width=\textwidth]{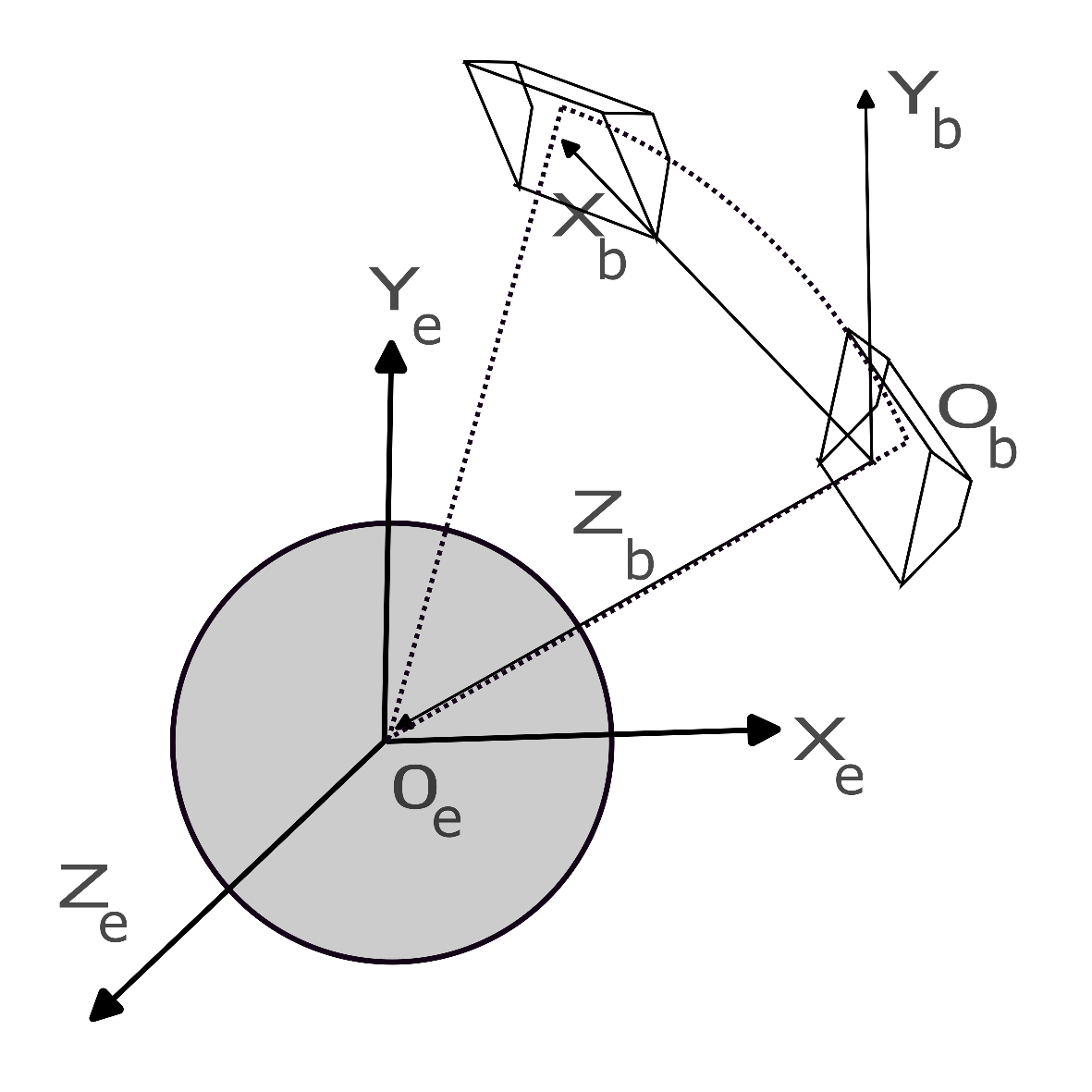}
         \caption{Coordinate system}
         \label{fig:coordinate_system}
     \end{subfigure}
     \hfill
     \begin{subfigure}[b]{0.3\textwidth}
         \centering
         \includegraphics[width=0.9\textwidth]{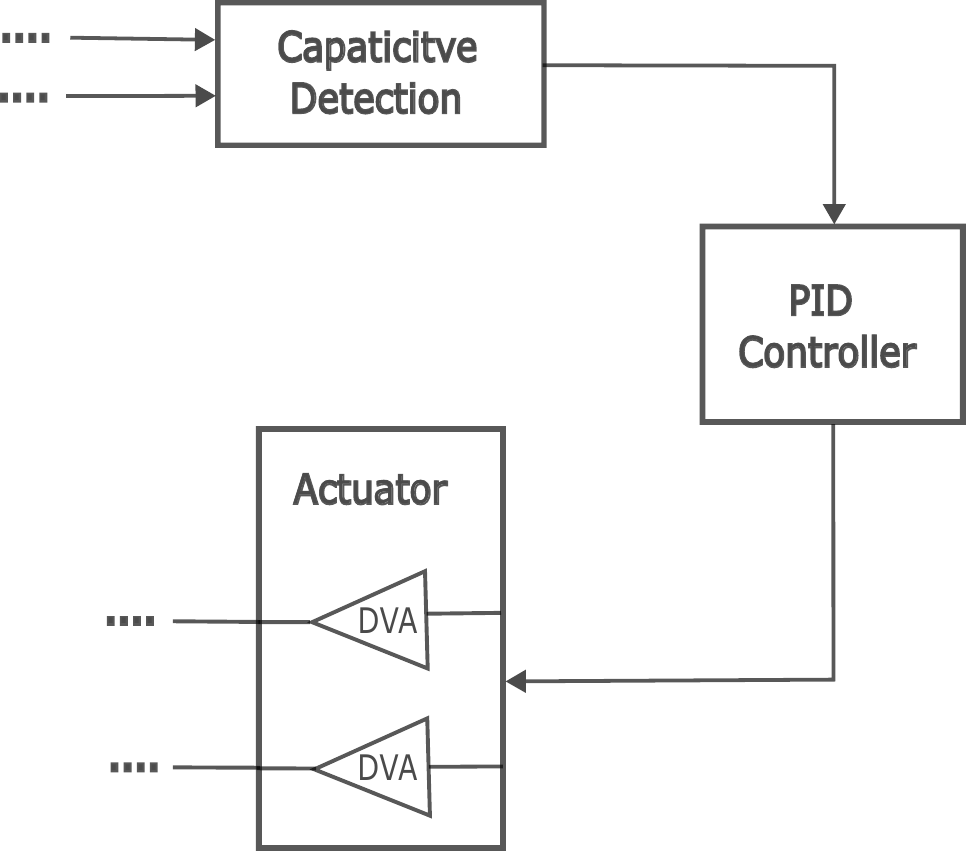}
         \caption{PID controller \cite{kornfeld2019grace_grace_2}}
         \label{fig:pid_controller}
     \end{subfigure}
     \hfill
     \begin{subfigure}[b]{0.3\textwidth}
         \centering
         \includegraphics[width=\textwidth]{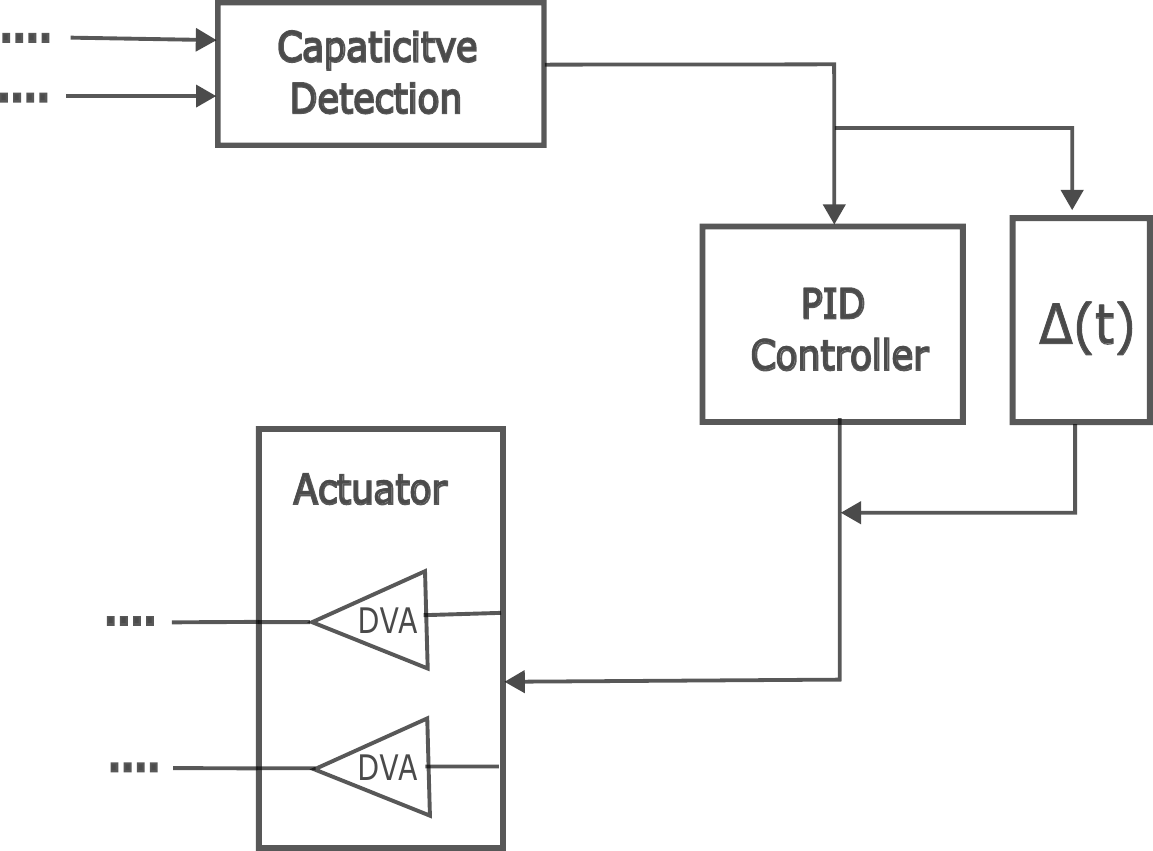}
         \caption{Proposed minimally modified PID controller}
         \label{fig:modified_pid_controller}
     \end{subfigure}
        \caption{(a) The coordinate system. Figs. (b) and (c) represent the actual attitude PID control block in GRACE-FO satellites \cite{kornfeld2019grace_grace_2} and the proposed minimally modified PID controller, respectively, with GRU based time varying disturbance compensator $\Delta(t)$. }
        \label{}
\end{figure}

In this paper, we use the following right-handed Cartesian coordinate system. First, the origin of the Inertial Frame (IF) $ O X_e Y_e Z_e $ is at the Earth's center of mass. As shown in Fig. \ref{fig:coordinate_system}, the $ O Z_e $ axis aligns with the Earth's rotation axis, and the $ O X_e $ axis points towards the vernal equinox of the J2000 epoch. $OY_e$ is obtained from the right-hand rule. 

Second, the Body Frame (BF) $ O_b X_b Y_b Z_b $. The axes of the BF correspond to the satellite's principal axes of inertia, with $ O_b X_b $ aligning with the line of sight of the ranging device, and $ O_b Z_b $ orthogonal to the satellite's bottom. Finally, the Reference Frame (RF) $ O_r X_r Y_r Z_r $ is the frame that defines the desired attitude of the satellite with respect to the other satellite.

First, we construct the reference motion for the second satellite, which is almost identical to that of the first. The basis vectors of the RF are defined as follows; $ \boldsymbol{e}_1 $ aligns with the line of sight between the two satellites, $ \boldsymbol{e}_2 $ is orthogonal to both the line of sight and the satellite's radius vector, and $ \boldsymbol{e}_3 $ completes the system as a right orthogonal vector. Let $ \boldsymbol{r}_1, \boldsymbol{r}_2$ and $\boldsymbol{v}_1, \boldsymbol{v}_2 $ represent the positions and velocities of the first and second satellites, respectively, which are assumed to be known a priori. Then, the basis vectors are defined as:
\begin{align*}
\boldsymbol{e}_1 = \frac{\boldsymbol{r}_1 - \boldsymbol{r}_2}{\|\boldsymbol{r}_1 - \boldsymbol{r}_2\|},\;\;\;\; \boldsymbol{e}_3 = -\frac{\boldsymbol{r}_2 - \boldsymbol{e}_1 \langle \boldsymbol{r}_2, \boldsymbol{e}_1\rangle}{\|\boldsymbol{r}_2 - \boldsymbol{e}_1 \langle \boldsymbol{r}_2, \boldsymbol{e}_1 \rangle \|},\;\;\;\; \boldsymbol{e}_2 = \boldsymbol{e}_3 \times \boldsymbol{e}_1 .
\end{align*}
where $ \langle\cdot, \cdot\rangle $ denotes the dot product, and $ \times $ denotes the cross product between two vectors.

\subsection{Nonlinear attitude dynamics}

The kinematics and the dynamics of the spacecraft attitude system are modeled as
\begin{align}
    \dot{R}=R\boldsymbol{\omega},\quad\quad I\dot{\bomega}=-\bomega\times(I\bomega)+\boldsymbol{\tau}+\bd
    \label{eqn:attitude_dynamics}
\end{align}
where $\bomega\in\mathbb{R}^3$ is the angular velocity of the spacecraft, $R\in SO(3)$ ($SO(3)$ denotes the 3D rotation group) is the rotation matrix, $I\in\mathbb{S}^3_+$ ($\mathbb{S}^3_+$ denotes the set of symmetric positive definite $3\times 3$ matrices) is the inertia, $\boldsymbol{\tau}\in\mathbb{R}^3$ is the control torque and $\bd \in\mathbb{R}^3$ is the time-varying net external disturbances acting on the spacecraft. Note that the attitude and the angular rate are with respect to the reference frame (RF) that is estimated from the ground-based station. We assume that this RF is known precisely for our simulations. Furthermore, the main objective of this paper is two-fold (Section \ref{subsection:proposed_approach}). The first is to estimate the disturbances $\bd$, given the time series data of the relative attitude and angular rates. Second, is to modify the traditionally PID controller minimally so as to actively compensate for the additive external disturbances acting on the satellites.

\subsection{Feedforward neural network (FNN)}

\begin{figure}[htbp]
    \centering
    \begin{subfigure}[b]{0.5\textwidth}
        \centering
        \resizebox{0.7\textwidth}{!}{
        \begin{tikzpicture}[x = 3.5cm,y = 1.4cm, scale = 0.7]
  \message{^^JNeural network without arrows}
  \readlist\Nnod{3,7,7,2} 
  
  \message{^^J  Layer}
  \foreachitem \N \in \Nnod{
    \def\lay{\Ncnt} 
    \pgfmathsetmacro\prev{int(\Ncnt-1)} 
    \message{\lay,}
    \foreach \i [evaluate={\y=\N/2-\i; \x=\lay; \n=\nstyle;}] in {1,...,\N}{ 
      
      \node[node \n] (N\lay-\i) at (\x,\y) {$n_\i^{(\prev)}$};
      
      \ifnum\lay>1 
        \foreach \j in {1,...,\Nnod[\prev]}{ 
          \draw[connect,white,line width=1.2] (N\prev-\j) -- (N\lay-\i);
          \draw[connect] (N\prev-\j) -- (N\lay-\i);
        }
      \fi 
      
    }
  }
  \node[above=0.25,align=center,myblue!60!black] at (N1-1.90) {Input\\[-0.2em]};
  \node[above=0.25,align=center,mypurple!60!black] at (2.5,3) {Hidden layers};
  \node[above=0.25,align=center,mygreen!60!black] at (N\Nnodlen-1.90) {Output\\[-0.2em]};
  
\end{tikzpicture}}
        \subcaption{Schematic of a neural network (NN)}
        \label{fig:nn}
    \end{subfigure}
    \hfill
    \begin{subfigure}[b]{0.45\textwidth}
        \centering
        \resizebox{0.7\textwidth}{!}{
        \begin{tikzpicture}[line width = 0.5mm]
\node[circle, draw, minimum size=2cm, line width=0.5mm, black, align=center] (projei) at (0,0) {%
  $\displaystyle\sum\limits_{i=1}^n w_i x_i +b$ \hspace{0.1cm}
  \hspace* {3pt} $\sigma(\cdot)$%
};

\draw[line width=0.5mm] ([yshift=-1.55cm,xshift=0.6cm]projei.center) -- ([yshift=1.55cm,xshift=0.6cm]projei.center);

\node[above left=1cm and 1cm of projei, label={left:$x_{1}$}] (inputi) {};
\node[below left=1cm and 1cm of projei, label={left:$x_{n}$}] (inputn) {};

\node at ($(inputi)!0.5!(inputn)$) {\vdots};

\node[right=3cm of projei] (outi) {};

\draw[->] (inputi) -- (projei);
\draw[->] (inputn) -- (projei);
\draw[->] (projei) -- node[above, midway] {$\sigma\left(\displaystyle\sum\limits_{i=1}^n w_i x_i+b\right)$} (outi);
\end{tikzpicture}}
        \subcaption{Schematic of a single neuron}
        \label{fig:neuron}
    \end{subfigure}
    \caption{Feedforward neural network (FNN)}
    \label{fig:fnn}
\end{figure}
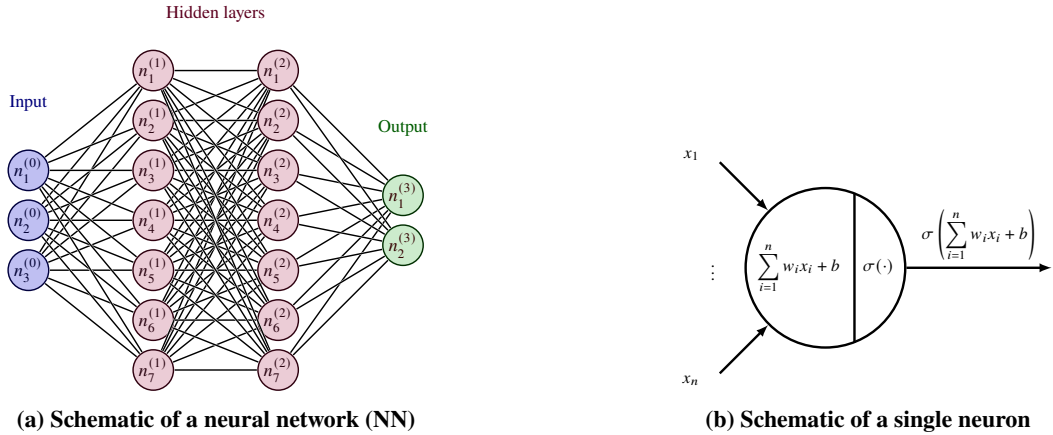

Neural networks (NN) are algorithms used for approximating functions based on data, structured through layers that start with the input and end with the output. These layers are organized into a hierarchy, where each level, except the first and last layers, is called a \textit{hidden} layer. By adjusting the number and size of these hidden layers, NNs can handle functions of different complexities. A typical NN configuration, featuring two hidden layers, is depicted in Fig. \ref{fig:nn}. Within this diagram, each circle symbolizes the basic computational element of NN, known as a neuron (Fig.\ref{fig:neuron}) which involves a linear transformation followed by a nonlinear operation.

A feedforward neural network (FNN) \cite{bebis1994feed} is essentially a NN characterized by a series of hidden layers. The term ``activation'' refers to the output generated by each layer. In an FNN, the activation output from a preceding layer becomes the input for the subsequent layer. Assuming $L$ is the total number of layers, the activation generated by the $j^{\text{th}}$ hidden layer can be recursively expressed as:
\begin{align*}
    \mathbf{z}^{(j)}=\sigma\left(W^{(j)} \mathbf{z}^{(j-1)}+\mathbf{b}^{(j)}\right), \forall j \in\{1,2, \cdots, L\},\quad \mathbf{z}^0=\boldsymbol{\xi},\quad d_0=D
    \label{eqn:fnn}
\end{align*}

where $\xi$ is the input vector, $W^{(j)}$ represents the weight matrix and $\mathbf{b}^{(j)}$ the bias vector for the $j^{\text{th}}$ layer, with $d_j$ indicating the number of neurons in that layer. Note that the activation function $\sigma$ denotes an element-wise applied nonlinear function, with common choices being the logistic, hyperbolic tangent, or ReLU functions \cite{goodfellow2016deep, dubey2022activation}.
The weight matrix $W^{(j)}$ and the biases $\mathbf{b}^{(j)}$ (for $j\in[1,L+1]_d$), are termed its parameters, collectively denoted as $\boldsymbol{\theta}=\left\{W^{(j)}, \mathbf{b}^{(j)}\right\}_{j=1}^{L+1} \in \boldsymbol{\Theta}$. These parameters, which consist of the weights and biases, fully describe the network structure.

\subsection{Recurrent neural network (RNN)}

\begin{figure}[htbp]
    \centering
    \begin{minipage}[c]{0.55\textwidth}
        \centering
        \resizebox{\textwidth}{!}{
        \begin{tikzpicture}[>=Latex, cell/.style={draw, rounded rectangle, minimum height=1cm, minimum width=1.5cm, fill=#1}, layer/.style={draw, rectangle, dashed, inner sep=5pt}]

\def\layerdistance{1cm} 
\def\largelayerdistance{2cm} 
\def\celldistance{1cm}
\def\largecelldistance{2cm}

\node[cell=yellow!20] (1-1) {\texttt{RNN Cell}};
\foreach \x [remember=\x as \lastx (initially 1)] in {2,3}{
    \node[cell=yellow!20, right=\celldistance of 1-\lastx] (1-\x) {\texttt{RNN Cell}};
}
\node[cell=yellow!20, right=\largecelldistance of 1-3] (1-4) {\texttt{RNN Cell}}; 

\foreach \x in {1,...,4}{
    \node[cell=blue!20, above=\layerdistance of 1-\x] (2-\x) {\texttt{RNN Cell}};
}

\foreach \x in {1,...,4}{
    \node[cell=green!20, above=\largelayerdistance of 2-\x] (K-\x) {\texttt{RNN Cell}};
}

\foreach \x [evaluate=\x as \nextx using int(\x+1)] in {1,2,3}{
    \ifnum\x=3 
        \draw[->, dashed] (1-\x) -- (1-\nextx);
        \draw[->, dashed] (2-\x) -- (2-\nextx);
        \draw[->, dashed] (K-\x) -- (K-\nextx);
    \else
        \draw[->] (1-\x) -- (1-\nextx);
        \draw[->] (2-\x) -- (2-\nextx);
        \draw[->] (K-\x) -- (K-\nextx);
    \fi
}

\foreach \x in {1,...,4}{
    \draw[->] (1-\x) -- (2-\x);
    \draw[->, dashed] (2-\x) -- (K-\x);
}

\node[left=0.3cm of 1-1, align=right] {$1^\text{st}$ Layer};
\node[left=0.3cm of 2-1, align=right] {$2^\text{nd}$ Layer};
\node[left=0.3cm of K-1, align=right] {$K^\text{th}$ Layer};

\node[below=0.3cm of 1-1, align=center] {\Large $u_{t-w}$};
\node[below=0.3cm of 1-2, align=center] {\Large $u_{t-w-1}$};
\node[below=0.3cm of 1-3, align=center] {\Large $u_{t-w-2}$};
\node[below=0.3cm of 1-4, align=center] {\Large $u_{t-1}$};

\draw[->] (K-4.north) -- ++(0,1) node[above] {\Large$\hat{y}$};

\end{tikzpicture}
        }
        \subcaption{Schematic of RNN (many to one) with $K$ layers}
        \label{fig:rnn_schematic}
    \end{minipage}
     \begin{minipage}[c]{0.4\textwidth}
        \centering
        \resizebox{0.6\textwidth}{!}{

        \begin{tikzpicture}[
    font=\sf \scriptsize,
    >=LaTeX,
    cell/.style={
        rectangle, 
        rounded corners=5mm, 
        draw,
        very thick,
        fill=black!10
        },
    operator/.style={
        circle,
        draw,
        inner sep=-0.5pt,
        minimum height =.2cm,
        },
    function/.style={
        ellipse,
        draw,
        inner sep=1pt
        },
    ct/.style={
        circle,
        draw,
        line width = .75pt,
        minimum width=1cm,
        inner sep=1pt,
        fill=purple!20,
        },
    gt/.style={
        rectangle,
        draw,
        minimum width=4mm,
        minimum height=3mm,
        inner sep=1pt,
        fill=green!30,
        },
    mylabel/.style={
        font=\scriptsize\sffamily
        },
    ArrowC1/.style={
        rounded corners=.25cm,
        thick,
        },
    ArrowC2/.style={
        rounded corners=.5cm,
        thick,
        },
    cell/.style={
        rectangle, 
        rounded corners=5mm, 
        draw,
        very thick,
        fill=black!5
        },
    operator/.style={
        circle,
        draw,
        inner sep=-0.1pt,
        minimum height =.3cm,
        },
    function/.style={
        ellipse,
        draw,
        inner sep=1pt
        },
    gt/.style={
        rectangle,
        draw,
        minimum width=4mm,
        minimum height=3mm,
        inner sep=1pt,
        fill=green!20,
        },
    scale=0.8,
    every node/.style={scale=0.8}
    ]

    \node [cell, minimum height =4cm, minimum width=6cm] at (0,0){} ;

    \node [gt] (ibox1) at (-2,-0.75) {$\sigma$};
    \node [gt] (ibox2) at (-1.5,-0.75) {$\sigma$};
    \node [gt, minimum width=1cm] (ibox3) at (-0.5,-0.75) {Tanh};
    \node [gt] (ibox4) at (0.5,-0.75) {$\sigma$};

    \node [operator] (mux1) at (-2,1.5) {$\times$};
    \node [operator] (add1) at (-0.5,1.5) {+};
    \node [operator] (mux2) at (-0.5,0) {$\times$};
    \node [operator] (mux3) at (1.5,0) {$\times$};
    \node [function] (func1) at (1.5,0.75) {Tanh};

    \node[ct, label={[mylabel]}] (c) at (-4,1.5) {$c_{t-1}$};
    \node[ct, label={[mylabel]}] (h) at (-4,-1.5) {$h_{t-1}$};
    \node[ct, label={[mylabel]}] (x) at (-2.5,-3) {$u_t$};

    \node[ct, label={[mylabel]}] (c2) at (4,1.5) {$c_t$};
    \node[ct, label={[mylabel]}] (h2) at (4,-1.5) {$h_t$};
    \node[ct, label={[mylabel]}] (x2) at (2.5,3) {$h_t$};
  
    \draw [->, ArrowC1] (c) -- (mux1) -- (add1) -- (c2);

    \draw [ArrowC2] (h) -| (ibox4);
    \draw [ArrowC1] (h -| ibox1)++(-0.5,0) -| (ibox1); 
    \draw [ArrowC1] (h -| ibox2)++(-0.5,0) -| (ibox2);
    \draw [ArrowC1] (h -| ibox3)++(-0.5,0) -| (ibox3);
    \draw [ArrowC1] (x) -- (x |- h)-| (ibox3);

    \draw [->, ArrowC2] (ibox1) -- (mux1);
    \draw [->, ArrowC2] (ibox2) |- (mux2);
    \draw [->, ArrowC2] (ibox3) -- (mux2);
    \draw [->, ArrowC2] (ibox4) |- (mux3);
    \draw [->, ArrowC2] (mux2) -- (add1);
    \draw [->, ArrowC1] (add1 -| func1)++(-0.5,0) -| (func1);
    \draw [->, ArrowC2] (func1) -- (mux3);

    \draw [->, ArrowC2] (mux3) |- (h2);
    \draw (c2 -| x2) ++(0,-0.1) coordinate (i1);
    \draw [-, ArrowC2] (h2 -| x2)++(-0.5,0) -| (i1);
    \draw [->, ArrowC2] (i1)++(0,0.2) -- (x2);
    
    \node at (0,4) {\huge\texttt{LSTM Cell}};

    \def\verticalShift{-9} 

    \node [cell, minimum height=4cm, minimum width=6cm] at (0,\verticalShift) {};

    \node [gt] (gibox1) at (-0.5,\verticalShift-0.5) {$\sigma$};
    \node [gt] (gibox2) at (1,\verticalShift-0.5) {$\sigma$};
    \node [gt, minimum width=1cm] (gibox3) at (2,\verticalShift-0.5) {Tanh};

    \node [operator] (gimux1) at (1,\verticalShift+1.5) {$\times$};
    \node [operator] (giadd1) at (2,\verticalShift+1.5) {+};
    \node [operator] (gimux2) at (-1.5,\verticalShift+0.5) {$\times$};
    \node [operator] (gimux3) at (2,\verticalShift+0.25) {$\times$};
    \node [function] (gifunc2) at (1,\verticalShift+0.75) {1-};

    \node[ct, label={[mylabel]}] (gic) at (-4,\verticalShift+1.5) {$h_{t-1}$};
    \node[ct, label={[mylabel]}] (gix) at (-2.5,\verticalShift-3) {$u_t$};

    \node[ct, label={[mylabel]}] (gic2) at (4,\verticalShift+1.5) {$h_t$};
    \node[ct, label={[mylabel]}] (gix2) at (2.5,\verticalShift+3) {$h_t$};

    \draw [->, ArrowC1] (gifunc2) -- (gimux1);
    \draw [-, ArrowC1] (gibox2) -- (gifunc2);
    \draw [-, ArrowC1] (gibox3) -- (gimux3);
    \draw [->, ArrowC1] (gimux3) -- (giadd1);
    \draw [-, ArrowC1] (gimux1) -- (giadd1);
    \draw [-, ArrowC2] (gic) -- (gimux1);
    \draw [->, ArrowC2] (giadd1) -- (gic2);
    
    \coordinate (gA) at (2.5,\verticalShift+1.5);
    \draw [<-, ArrowC2] (gix2) -- (gA);
    
    \draw [->, ArrowC2] (gibox1) |- (gimux2);
    
    \coordinate (gB) at (-1.5,\verticalShift+1.5);
    \draw [-, ArrowC2] (gimux2) -- (gB);
    
    \coordinate (gC) at (-2.5,\verticalShift+1.5);
    \draw [-, ArrowC2] (gix) -- (gC);
    
    \coordinate (gD) at (-2.5,\verticalShift-1.5);
    \draw [-, ArrowC2] (gD) -- (2,\verticalShift-1.5) -- (gibox3);
    
    \coordinate (gE) at (-2.5,\verticalShift-1);
    \draw [-, ArrowC1] (gE) -- (1,\verticalShift-1) -- (gibox2);
    
    \coordinate (gF) at (-0.5,\verticalShift-1);
    \draw [-, ArrowC2] (gF) -- (gibox1);
    
    \coordinate (gG) at (-1.5,\verticalShift-0.9);
    \draw [-, ArrowC2] (gimux2) -- (gG);
    
    \coordinate (gH) at (-1.5,\verticalShift-1.1);
    \coordinate (gI) at (-1.5,\verticalShift-1.5);
    \draw [-, ArrowC2] (gH) -- (gI);
    
    \coordinate (gJ) at (1,\verticalShift+0.25);
    \draw [->, ArrowC2] (gJ) -- (gimux3);
    
    \node[label=above:{\fontsize{8}{12}\selectfont $r_t$}] at (-0.75,\verticalShift-0.5) {};
    \node[label=above:{\fontsize{8}{12}\selectfont $z_t$}] at (0.75,\verticalShift-0.5) {};
    \node[label=above:{\fontsize{8}{12}\selectfont $\tilde{h}_t$}] at (2.25,\verticalShift-0.5) {};

    \node at (0,-5) {\huge\texttt{GRU Cell}};

    \end{tikzpicture}
        
        }
        \subcaption{Popular variants of RNN cell}
        \label{fig:rnn_variants}
    \end{minipage}
    \caption{Recurrent neural network (RNN)}
    \label{fig:rnn}
\end{figure}
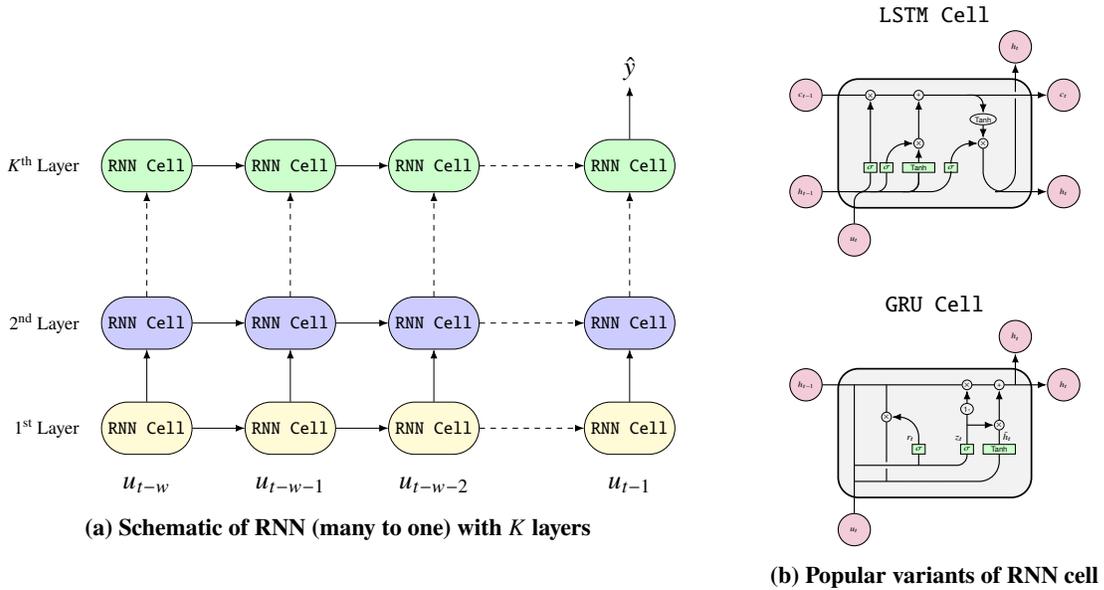

RNNs \cite{schmidt2019recurrent} are a class of artificial neural networks in which connections between neurons form a directed graph along a temporal sequence. 
This allows them to exhibit temporal dynamic behavior and process sequences of inputs, unlike FNN which is a static input-output mapping where the flow of information is unidirectional from input to output layers without any cycles.
The core of an RNN is its cell, which processes one input at a time in a sequence, maintaining a hidden state $\mathbf{h}_t$ that captures information about the past elements of the sequence. Assuming $\mathbf{u}_t$, $\mathbf{h}_t$, $\mathbf{y}_t$ represent the input, hidden state, and output at time $t$ respectively, the governing equations of a conventional RNN cell can be described as follows: 
\begin{equation*}
\mathbf{h}_t=\sigma\left(W_{hh} \mathbf{h}_{t-1}+W_{uh} \mathbf{u}_t+\mathbf{b}_h\right),\quad \quad \mathbf{y}_t=W_{h y} \mathbf{h}_t+\mathbf{b}_{y},
\label{eqn:rnn}
\end{equation*}
where $W_{hh}, W_{uh}$, and $W_{hy}$ represent the weight matrices, $\mathbf{b}_h$ and $\mathbf{b}_y$ are bias vectors, and $\sigma$ being a nonlinear activation function. RNN can predict an output given a window of inputs by processing each element of the input sequence one at a time, updating its hidden state based on the current input and the previous hidden state, as shown in Fig. \ref{fig:rnn_schematic}. This allows the network to make predictions based on the information it has seen up to the last $w$ time steps (Fig. \ref{fig:rnn}) of a sequence. LSTM and GRU are two popular RNN variants designed to address the problem of vanishing gradients \cite{bengio1994learning, pascanu2013difficulty} (which occurs in neural networks when the gradients of the loss function become exceedingly small during backpropagation, causing the weights in the previous layers to update minimally and hindering effective learning) in conventional RNNs, enabling them to learn long-range dependencies more effectively.

\subsubsection{Long short term memory (LSTM)}

LSTM networks \cite{hochreiter1997long} are a specialized form of RNN designed for the prediction of sequential data. The core of LSTM is a memory cell that maintains its state over time, managed by three gates: input ($\mathbf{i}_t$), forget ($\mathbf{f}_t$), and output ($\mathbf{o}_t$). These gates, along with a candidate value, regulate the flow and updating of information within the cell, effectively addressing the vanishing gradient problem seen in conventional RNNs. By selectively adding or removing information, LSTMs can preserve relevant information over longer time periods, significantly improving the flow and retention of information compared to traditional RNNs. 
Furthermore, the forget gate determines the extent of information to be discarded from the cell state (Fig. \ref{fig:rnn_variants}):
\begin{equation*}
    \mathbf{f}_{t} = \sigma(W_{u_{f}}\mathbf{u}_{t} + W_{h_{f}}\mathbf{h}_{t-1} + \mathbf{b}_{f}),
\end{equation*}
where $\mathbf{f}_t$ denotes the output of the forget gate at time $t$, $W_{u_{f}}$ and $W_{h_{f}}$ are the input and recurrent weights of the forget gate, $\mathbf{b}_{f}$ is the bias, and $\sigma(.)$ represents the sigmoid activation function. Following, the decision of the input gate to update the state is captured by:
\begin{equation*}
   \mathbf{i}_{t} = \sigma(W_{u_{i}}\mathbf{u}_{t} + W_{h_{i}}\mathbf{h}_{t-1} + \mathbf{b}_{i}). \label{input_gate}
\end{equation*}
A hyperbolic tangent function then generates a vector of candidate values $\mathbf{z}_t$ for state addition:
\begin{equation*}
    \mathbf{z}_{t} = \texttt{tanh}(W_{u_{z}}\mathbf{u}_{t} + W_{h_{z}}\mathbf{h}_{t-1} + \mathbf{b}_{z}),
\end{equation*}
with ($W_{u_{i}}, W_{h_{i}}, \mathbf{b}_{i}$) and ($W_{u_{z}}, W_{h_{z}}, \mathbf{b}_{z}$) denoting the weights and biases associated with the input gate and cell update mechanism, respectively. The cell state update is formulated as follows:
\begin{equation*}
    \mathbf{c}_{t} = \mathbf{f}_{t} \star  \mathbf{c}_{t-1} + \mathbf{i}_{t} \star  \mathbf{z}_{t},
\end{equation*}
where $\star $ signifies element-wise multiplication. Lastly, based on the output gate, certain parts of the cell state are outputted. The output of LSTM cell $\mathbf{h}_t$ is then calculated by applying a hyperbolic tangent function to the cell state and multiplying the result by the output gate, as shown below:
\begin{equation*}
     \label{output_gate}
     \begin{split}
     &\mathbf{o}_{t} = \sigma(W_{u_{o}}\mathbf{u}_{t} + W_{h_{o}}\mathbf{h}_{t-1} + \mathbf{b}_{o}),\quad\quad \mathbf{h}_{t} = \mathbf{o}_{t} \star  \texttt{tanh}(\mathbf{c}_{t}),
     \end{split}
 \end{equation*}
where ($W_{u_{o}}, W_{h_{o}}, \mathbf{b}_{o}$) represents the recurrent weights, input weights, and bias of the output gate respectively.

\subsubsection{Gated recurrent unit (GRU)}
Gated Recurrent Units (GRUs) are a streamlined variant of LSTM networks, introduced in \cite{cho2014learning} to simplify the LSTM architecture by eliminating the cell state and reducing the number of gates from three to two: the update gate ($\mathbf{z}_t$) and the reset gate ($\boldsymbol{r}_t$). These modifications allow for fewer computations during training, improving efficiency without compromising predictive performance. The update gate determines the extent of information transition from the previous state, while the reset gate integrates the current and previous states, adjusting the influence of past information. Fig. \ref{fig:rnn_variants} illustrates the structural design of a GRU cell.
 
The update law for GRU is given by:
\begin{equation*}
    \mathbf{z}_{t}=\sigma\left(W_{u_{z}}\mathbf{u}_{t} + W_{h_{z}}\mathbf{h}_{t-1} + \mathbf{b}_{z} \right).
\end{equation*}
This update gate $\mathbf{z}_t$ helps in deciding the extent to which information from past time steps is forwarded into the future time steps. Consequently, the reset gate $\boldsymbol{r}_t$ is computed, which is important in determining the level of past information to forget. Mathematically,
\begin{equation*}
    \boldsymbol{r}_{t}=\sigma\left(W_{u_{r}}\mathbf{u}_{t} + W_{h_{r}}\mathbf{h}_{t-1} + \mathbf{b}_{r}\right), 
\end{equation*}
where $W_{u_{r}}$, $W_{h_{r}}$, and $\mathbf{b}_{r}$ denote the input weights, recurrent weights, and bias associated with the reset gate, respectively. Subsequently, the GRU formulates a new memory content, leveraging the reset gate to retain pertinent past information, expressed as:
\begin{equation*}
    \mathbf{\tilde{h}}_{t}=\texttt{tanh} \left(W_{u_{h}}\mathbf{u}_{t} + W_{h_{h}}\mathbf{h}_{t-1} + \mathbf{b}_{h}\right),
\end{equation*}
where the $\texttt{tanh}$ is the hyperbolic tangent function, and $W_{u_{h}}$, $W_{h_{h}}$, and $\mathbf{b}_{h}$ are the input weights, recurrent weights, and bias for the new memory content $\tilde{h}_{t}$ respectively. Finally, the update law for the current hidden state $\mathbf{h}_t$ is given by
\begin{equation*}
    \mathbf{h}_{t}=\left(1-\mathbf{z}_{t}\right)\star \mathbf{h}_{t-1} + \mathbf{z}_{t}\star \mathbf{\tilde{h}}_{t},
\end{equation*}
with $\star $ denoting the element-wise (Hadamard) multiplication.

\subsection{Attitude and orbit control system\label{subsec:control_system}}

The control torque $\boldsymbol{\tau}$ generated by the magnetorquers from the GRACE-FO satellites is given by \cite{kinzie2023dual_grace_fo}
\begin{align}
\boldsymbol{\tau} = \mathbf{m} \times \mathbf{B},
\end{align}
where $\mathbf{m}$ is the magnetic dipole vector created by three perpendicular magnetorquers, and $\mathbf{B}$ is the external magnetic field (for instance earth's magnetic field and due to other external sources which can be measured). Since the control torque is always orthogonal to the magnetic field's direction, there is always a direction at any given moment where the control torque cannot be generated. Various methods exist for using magnetorquers for orbital or inertial stabilization, most of which are based on PID controllers \cite{celani2015robust_pd_3,ovchinnikov2015three_pd_2,lovera2005global_pd_1}. In these methods, the dipole vector $\mathbf{m}$ is selected as follows:
\begin{align}
\mathbf{m} = \mathbf{B} \times \underbrace{\left(-k_p \boldsymbol{e} - k_d \dot{\boldsymbol{e}}-k_i\int_0^t\boldsymbol{e}(s) \mathrm{d}s\right)}_{\boldsymbol{u}_{\text{PID}}}
\label{eqn:pid_controller}
\end{align}
where $\boldsymbol{e}$ is the relative attitude error between the two GRACE-FO satellites, $k_p>0$, $ k_d>0$, and $k_i$ are the PID gains. Consequently, when the magnetic field is parallel to the applied control input (which usually occurs near the equator), the net dipole moment $\mathbf{m}$ becomes zero, resulting in zero net torque. When this occurs, the torque for the satellites is generated using the thrusters present on the satellites.

\subsection{Proposed iterative approach\label{subsection:proposed_approach}}
In this section, we present our proposed approach that minimally modifies the traditional PID controller $\boldsymbol{u}_{\text{PID}}$ to mitigate the relative attitude error (or improve the pointing accuracy) between the two satellites. We assume that the post-facto ground processed relative attitude estimate (or the desired targeted pointing) between the satellites is known a priori. This ground-processed relative attitude estimate is used for computing the actual external disturbances acting on the satellites and compensating them via a modified PID controller as discussed in the following. The proposed approach consists mainly of two steps. In the first step, the time series data of the relative attitude error from the GRACE-FO data product (for a particular time period, say $T>0$) are used to estimate the additive disturbances acting on the GRACE-FO satellites. The disturbance values are then fed to the GRU network to predict the disturbances for the next time period $T$. Using these predictions, $\boldsymbol{u}_{\text{PID}}$ is modified as follows:
\begin{align*}
\boldsymbol{u}^1_{\text{PID}}=\boldsymbol{u}_{\text{PID}}-\Delta_1(t)\quad\forall\;\; t\in[T,\;2T],
\label{eqn:pid_first_iteration}
\end{align*}
where the superscript in $\boldsymbol{u}^1_{\text{PID}}$ denotes the first iteration (or first modification) in the original $\boldsymbol{u}_{\text{PID}}$ and $\Delta_1(t)$ is the GRU based prediction for the disturbances for the time period $T$. Substituting the updated PID controller in \eqref{eqn:attitude_dynamics} gives 
\begin{align*}
       \dot{R}=R\bomega,\quad\quad I\dot{\bomega}=-\bomega\times(I\bomega)+\boldsymbol{u}_{\text{PID}}+\underbrace{\bd-\Delta_1(t)}_{\bd_1} \quad\forall\;\; t\in[T,\;2T].
\end{align*}
The additional $\Delta_1$ (from the trained GRU network using estimated disturbance data from time $[0,\;T]$) term tries to compensate for the disturbances $\bd$ as much as possible). Let the virtual disturbances that are acting on the satellites be given by $\bd_1:=\bd-\Delta_1(t)$.
These two steps are repeated again. In particular, using the modified PID controller $\boldsymbol{u}^1_{\text{PID}}$ \eqref{eqn:pid_first_iteration}, the states of the satellites are propagated for a time period $T$ i.e. for $t\in[T,\;2T]$. Subsequently, these states are used to estimate the virtual disturbances (i.e., $\bd_1$ for the first iteration) and thereafter update the trained GRU model. Note that the GRU model must be updated as the disturbance values $\bd$ for the time period $t\in[0,\;T]$ might be different from the disturbance values during the time period $t\in[T,\;2T]$. Furthermore, $\bd_1$ is termed as the virtual disturbance because they are the unknown residue obtained from subtracting the GRU predicted values and the actual disturbances from the previous iteration. Consequently,  $\boldsymbol{u}^N_{\text{PID}}$ after $N$ iterations can be written as follows:
\begin{align}
\boldsymbol{u}^N_{\text{PID}}=\boldsymbol{u}_{\text{PID}}-\sum_{i=1}^N\Delta_i(t),\quad\forall\;\; t\in[NT,\;(N+1)T].
\end{align}
The dynamics are then given by
\begin{align*}
       \dot{R}=R\bomega,\quad\quad I\dot{\bomega}=-\bomega\times(I\bomega)+\boldsymbol{u}_{\text{PID}}+\underbrace{\bd-\sum_{i=1}^N\Delta_i(t)}_{\bd_N},\quad\forall\;\; t\in[NT,\;(N+1)T],
\end{align*}
where $\Delta_k$ is the trained GRU model trained over the time series data of $\bd-\sum_{i=1}^{k-1}\Delta_i(t)$ (virtual disturbances) where $k=\{1,\dots,N\}$. Algorithm \ref{alg:proposed_approach} summarizes the proposed approach. The training methodology for the GRU is described in Algorithm \ref{alg:gru_bptt}. Given the susceptibility of neural networks to overfitting, we use early stopping as a regularization strategy, characterized by patience of $\textit{p}$ epochs. This implies that the training process is terminated if there is no improvement in loss over a consecutive span of $\textit{p}$ epochs.
Note that depending on the computational bandwidth of the satellites, it is always possible to increase/decrease the time period $T$ to estimate (from the relative error in the previous iteration) and predict disturbances (for the next iteration).
\begin{algorithm}[htbp]
\caption{Iterative approach}
\begin{algorithmic}[1]
    \State Initialize the relative attitude error $\mathcal{A}_0$ from the GRACE-FO data product and set $i=0$
    \State Define $\bu^0_{\text{PID}}:=\bu_{\text{PID}}$ as PID controller \eqref{eqn:pid_controller}
    \Repeat
        \State Extract external disturbance time series data obtained from $\mathcal{A}_i$, attitude dynamics \eqref{eqn:attitude_dynamics}, and $\bu^i_{\text{PID}}$
        \State Store these disturbances in set $\mathcal{D}^e_i$
        \State Learn time series trend in data $\mathcal{D}^e_i$ using Algorithm \ref{alg:gru_bptt}
        \State Modify control input: $\bu^{i+1}_{\text{PID}} \leftarrow \bu^i_{\text{PID}} - \Delta_i(t)$
        \Comment{$\Delta_i(t)$ is a GRU based model}
        \State Propagate the attitude dynamics \eqref{eqn:attitude_dynamics} using the modified PID controller $\bu^{i+1}_{\text{PID}}$ for time $t\in[(i+1)T,\;(i+2)T]$
        \State $i\leftarrow i+1$
    \Until{a stopping criterion is met}
\end{algorithmic}
\label{alg:proposed_approach}
\end{algorithm}

\begin{algorithm}[htbp]
\caption{Training GRU network via backpropagation} 
\label{alg:gru_bptt}
\begin{algorithmic}[1]
    \State Dataset $\mathcal{D}^e_i = \{d_1, d_2, ..., d_T\}$
    \State \textbf{Parameters}: Loss function $L(\theta)$, $\theta$: parameters of the GRU network, initial parameter $\theta_0$, learning rate $\alpha$, batch size $b$, number of epochs $E$, input window size $w$, number of layers $L$, number of hidden units in each layer $h$,  patience $p$. 
    \State Initialize best loss $L_{\text{best}} \approx \infty$
    \State Initialize patience counter $p_{\text{counter}} = 0$
      \For{$i=1 \ldots$ $E$} 
        \State $(d_t,\{d_{t-k}, ..., d_{t-1}\}) =$ Sample training batch
       \State $\theta:=\theta_{0} \sim p\left(\theta_{0}\right)$ \Comment{Initialize the weights from a Glorot uniform distribution} 
            \For{$j=1 \ldots b$}
                \State $\hat{d}_{t} = \texttt{GRU}\left(\{d_{t-w}, ..., d_{t-1}\};\theta, L, h\right)$
                \State $L_{\mathrm{total}}=\sum_{j=1}^{b} L\left(d_{j}, \hat{d}_{j}\right)$\Comment{$L$ is Huber loss between the actual and the GRU based predicted disturbances}
                    \State $\nabla L(\theta)=\frac{\partial L_{\text {total }}}{\partial \theta}$;
                \State $\theta=\theta-\alpha \nabla L(\theta)$\Comment{Adam optimizer for updating parameters $\theta$}
                \EndFor
       \If{$L_{\mathrm{total}} < L_{\text{best}}$}
            \State $L_{\text{best}} = L_{\mathrm{total}}$
            \State $p_{\text{counter}} = 0$
        \Else
            \State $p_{\text{counter}} = p_{\text{counter}} + 1$
        \EndIf
        \If{$p_{\text{counter}} > p$}
            \State \textbf{break}
        \EndIf
    \EndFor
    \State \Return $\theta$
\end{algorithmic} 
\end{algorithm}

\section{Results\label{sec:results}}
In this section, we discuss the efficacy of the proposed modified PID controller in attenuating the effects of external disturbances on the GRACE satellites. The implementation of the training and testing phases of this approach is programmed using the $\texttt{equinox}$ and $\texttt{optax}$ library with a JAX \cite{jax2018github} backend.
The advantages of using JAX over other popular open source machine learning libraries, such as PyTorch \cite{paszke2019pytorch} and TensorFlow \cite{abadi2016tensorflow}, lie in its functional programming paradigm and automatic differentiation capabilities, which are particularly advantageous for high-performance computation tasks and real-time implementability (10-100x speedup compared to PyTorch or TensorFlow).
The code for replicating the results of this paper is available from \href{https://github.com/shrenikvz/gru-attitude-control-geodetic-missions}{https://github.com/shrenikvz/gru-attitude-control-geodetic-missions}. 

\begin{table}[H]
\begin{minipage}{0.5\textwidth}
\centering
\caption{Tuned hyperparameters for GRU network}
\label{tab:hyperparameters}
\resizebox{\textwidth}{!}{%
\begin{tabular}{l r}
\hline
\textbf{Sampling Time}                   & 1  s                                                     \\ \hline
\textbf{No. of Layers}                   & 3                                                           \\ \hline
\textbf{No. of hidden units}             & 128                          \\ \hline
\textbf{Loss Function}                   & Huber                                                       \\ \hline
\textbf{Learning Rate}                   & 0.005                      \\ \hline
\textbf{Regularization}                  & Early Stop. of Loss (Pat. 50)                        \\ \hline
\textbf{Optimizer}                       & Adam ($\beta_{1} = 0.9$, $\beta_{2} = 0.999$)               \\ \hline
\textbf{Weight Initialisation}           & Glorot Uniform                                              \\ \hline
\textbf{Batch Size}           & 64                                         \\ \hline
\textbf{Window Size}                     & 5 s                                                         \\ \hline
\textbf{Max Epochs}                      & 500 \\ \hline
\textbf{No. of times trained}            & 5                                                           \\ \hline
\end{tabular}}
\end{minipage}\hspace{0.05\textwidth}%
\begin{minipage}{0.4\textwidth}
\centering
\caption{Average computational training time for GRU network (when the network is trained 5 times)}
\label{tab:comp_time_gru}
\resizebox{0.5\textwidth}{!}{%
\begin{tabular}{|c|c|}
\hline
\textbf{Iterations $N$}                   & \textbf{Time}                                                    \\ \hline
1                  &  2.17 s                                                           \\ \hline
2             &  2.29 s                          \\ \hline
3                   &  1.31 s                                                       \\ \hline
4                  &  1.63 s                      \\ \hline
\end{tabular}
\label{tab:comp_time}  }
\end{minipage}
\end{table}

\begin{figure}[htbp]
    \centering
    \begin{subfigure}[b]{0.45\textwidth}
        \centering
        \includegraphics[width = 8.4 cm]{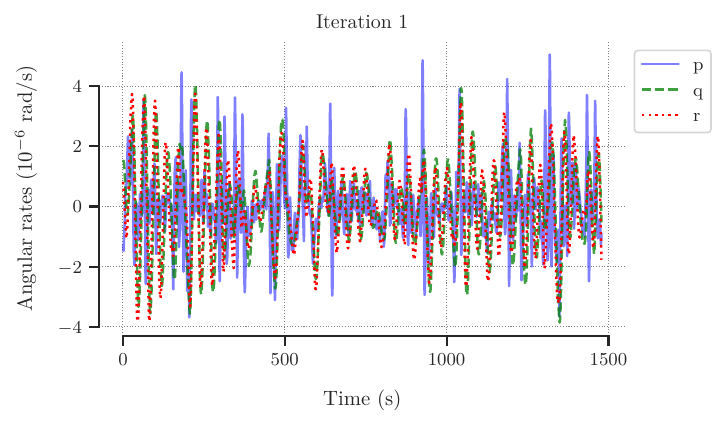}
    \end{subfigure}
    \hfill
    \begin{subfigure}[b]{0.45\textwidth}
        \centering
        \includegraphics[width = 8.4 cm]{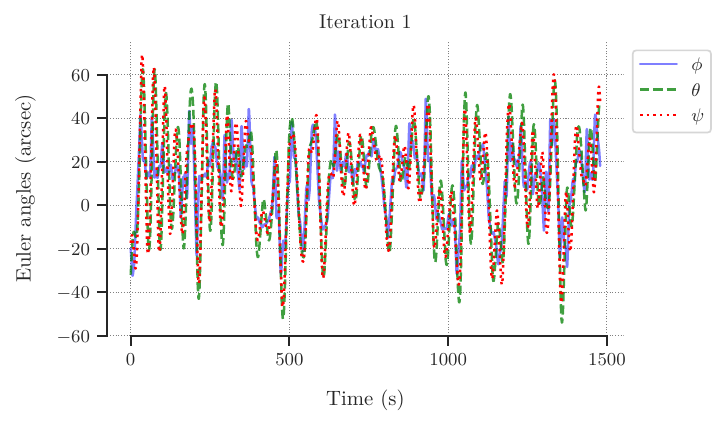}
    \end{subfigure}
    \begin{subfigure}[b]{0.45\textwidth}
        \centering
        \includegraphics[width = 8.4 cm]{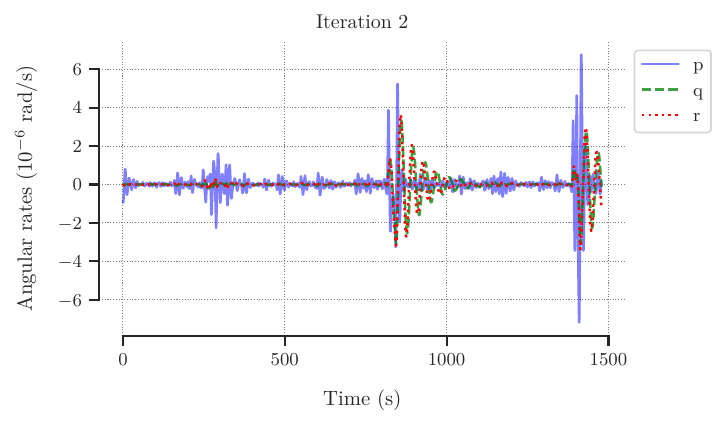}
    \end{subfigure}
    \hfill
    \begin{subfigure}[b]{0.45\textwidth}
        \centering
        \includegraphics[width = 8.4 cm]{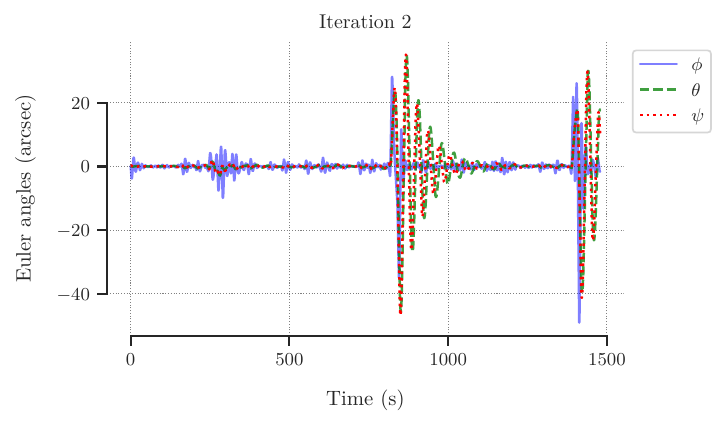}
    \end{subfigure}

    \begin{subfigure}[b]{0.45\textwidth}
        \centering
        \includegraphics[width = 8.4 cm]{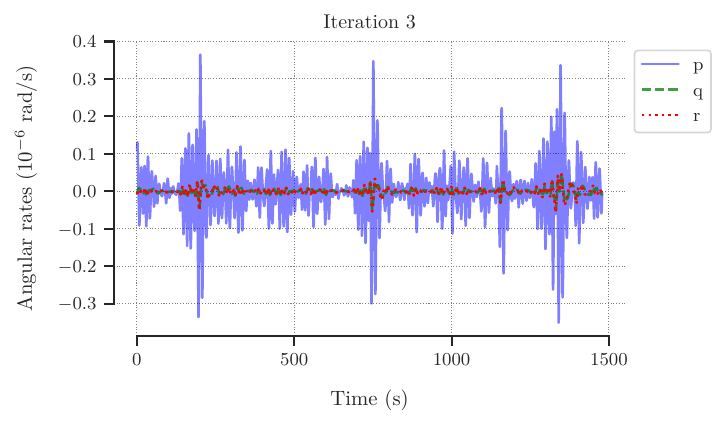}
    \end{subfigure}
    \hfill
    \begin{subfigure}[b]{0.45\textwidth}
        \centering
        \includegraphics[width = 8.4 cm]{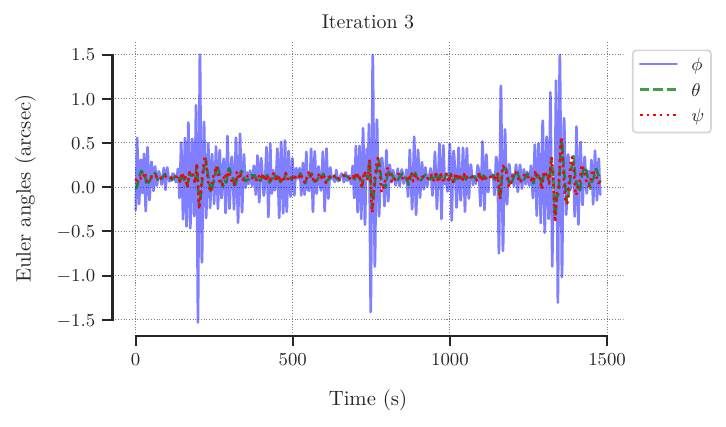}
    \end{subfigure}

    \begin{subfigure}[b]{0.45\textwidth}
        \centering
        \includegraphics[width = 8.4 cm]{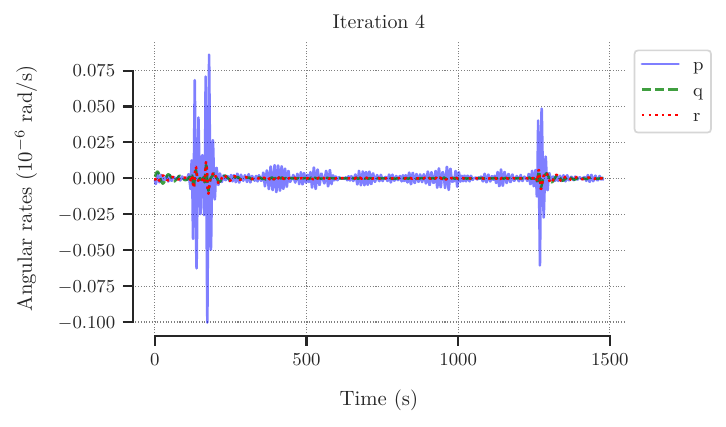}
    \end{subfigure}
    \hfill
    \begin{subfigure}[b]{0.45\textwidth}
        \centering
        \includegraphics[width = 8.4 cm]{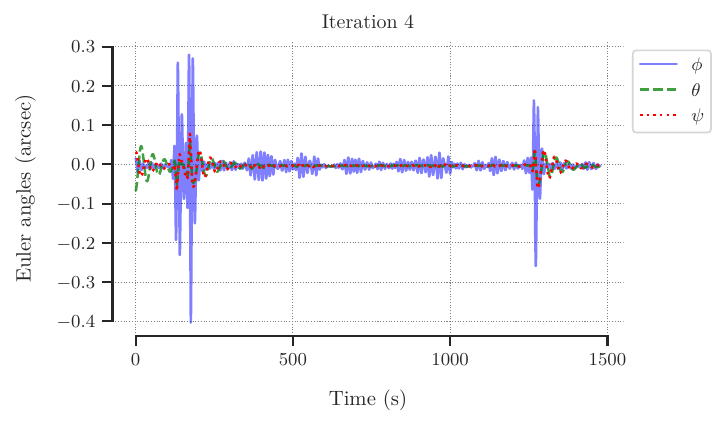}
    \end{subfigure}
    \caption{Evolution of angular rates and relative attitude with time across different iterations using the modified PID controller $\bu^i_{\text{PID}}$ ($i\in\{1,\dots,4\}$)}
    \label{fig:time_history}
\end{figure}

\begin{figure}[htbp]
    \centering
    \begin{subfigure}[b]{0.45\textwidth}
        \centering
        \includegraphics[width = 8.4 cm]{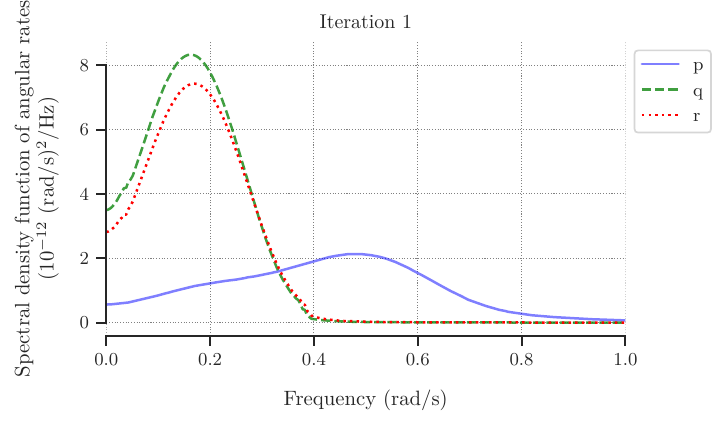}
    \end{subfigure}
    \hfill
    \begin{subfigure}[b]{0.45\textwidth}
        \centering
        \includegraphics[width = 8.4 cm]{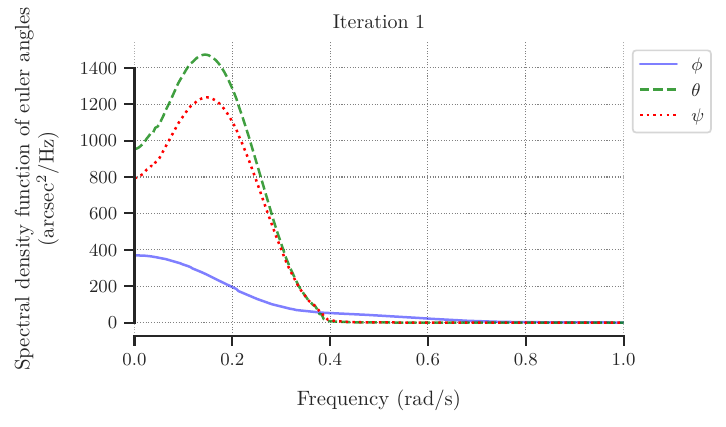}
    \end{subfigure}
    \begin{subfigure}[b]{0.45\textwidth}
        \centering
        \includegraphics[width = 8.4 cm]{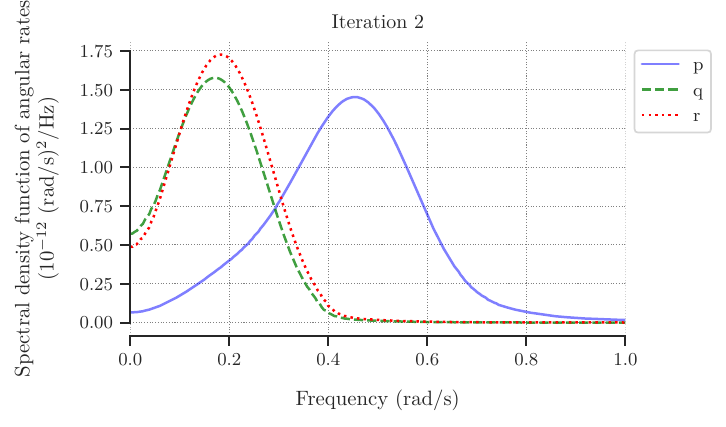}
    \end{subfigure}
    \hfill
    \begin{subfigure}[b]{0.45\textwidth}
        \centering
        \includegraphics[width = 8.4 cm]{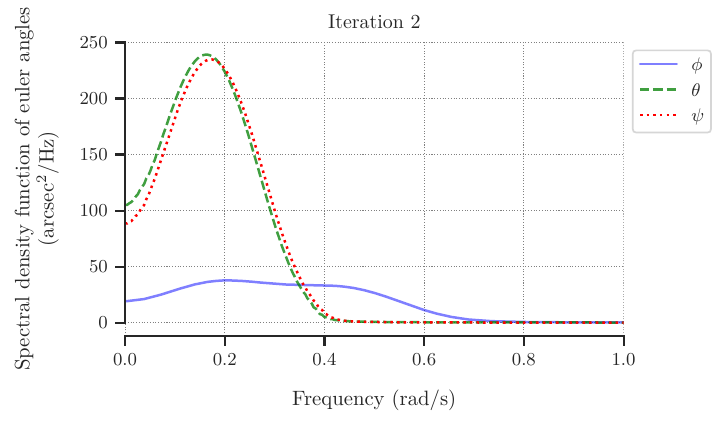}
    \end{subfigure}

    \begin{subfigure}[b]{0.45\textwidth}
        \centering
        \includegraphics[width = 8.4 cm]{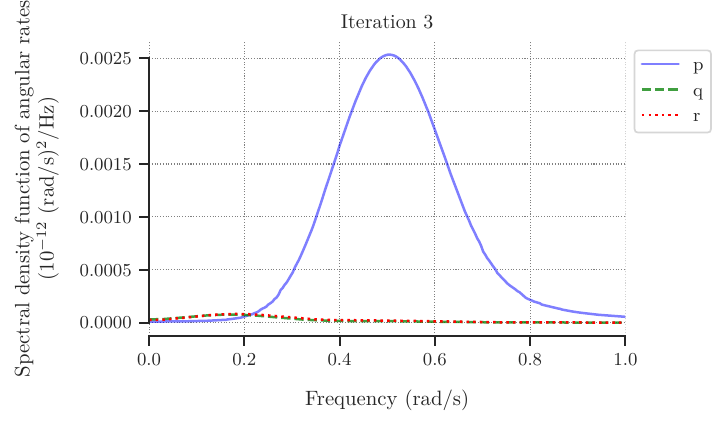}
    \end{subfigure}
    \hfill
    \begin{subfigure}[b]{0.45\textwidth}
        \centering
        \includegraphics[width = 8.4 cm]{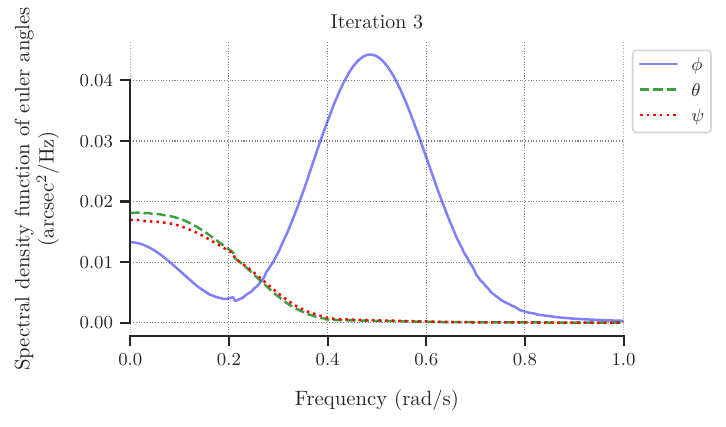}
    \end{subfigure}

    \begin{subfigure}[b]{0.45\textwidth}
        \centering
        \includegraphics[width = 8.4 cm]{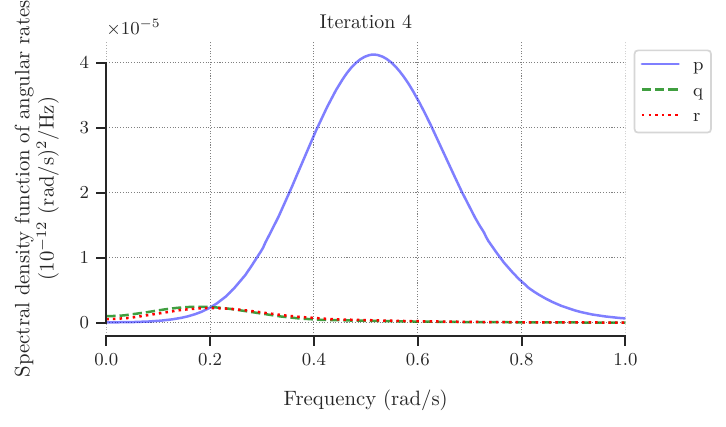}
    \end{subfigure}
    \hfill
    \begin{subfigure}[b]{0.45\textwidth}
        \centering
        \includegraphics[width = 8.4 cm]{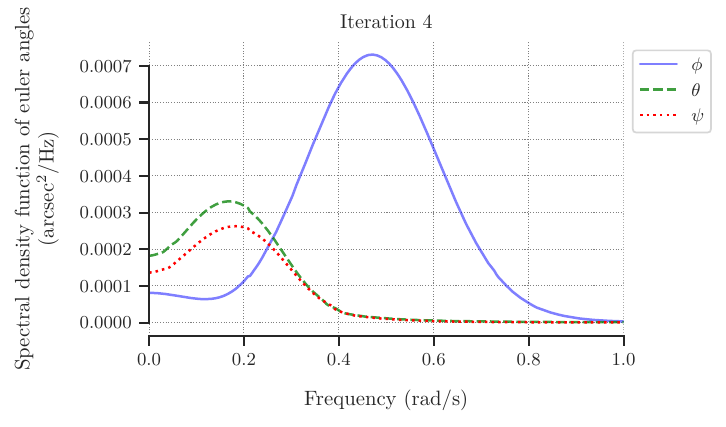}
    \end{subfigure}
    \caption{Spectrum plots for angular rates and relative attitude across different iterations using the modified PID controller $\bu^i_{\text{PID}}$ ($i\in\{1,\dots,4\}$)}
    \label{fig:psd}
\end{figure}

Note that the neural networks were trained five times for each iteration to ensure consistency and reliability in performance, thus mitigating issues associated with random initializations, hyperparameter sensitivity, and the stochastic nature of the training process.
In this work, a Huber loss is used because outliers in the training data are usually not handled properly by mean squared error (MSE). Furthermore, the mean absolute error (MAE) is computationally expensive as it uses the modulus operator function and may lead to local minima issues. The Huber loss is a hybrid of the MSE and MAE, operating as the MSE when the error is less than a specific threshold $\delta>0$, and as the MAE when the error exceeds this threshold. This is mathematically represented as follows:
\begin{equation*}
L_{\delta}(\boldsymbol{y}, \hat{\boldsymbol{y}})= \begin{cases}\frac{1}{2}(\boldsymbol{y}-\hat{\boldsymbol{y}})^{2} & \text{ for }\mid\boldsymbol{y}-\hat{\boldsymbol{y}})\mid \leq \delta \\ \delta \cdot\left(\mid \boldsymbol{y}-\hat{\boldsymbol{y}})\mid-\frac{1}{2} \delta\right) & \text{ otherwise}\end{cases}.
\end{equation*}
The above loss function is shown to be computationally efficient and is also able to handle outliers in the data. However, a notable limitation is that the hyperparameters need to be optimized to maximize model accuracy. In our algorithm, we use the Adam optimizer \cite{kingma2014adam}, with the hyperparameters tuned for optimal performance (see Table \ref{tab:hyperparameters}). 

The training has been performed on a Macbook Pro with an Apple M1 pro chip and 16 GB of memory. Table \ref{tab:comp_time} shows the average computational time (in s) to train the GRU network for different number of total iterations $N$. As seen from this table, the average computational time has an overall decreasing trend with $N$. This is mainly because, as $N$ increases, the amount of training data available to train the GRU network is less. Furthermore, it must be noted that depending on the computational budget on the GRACE-FO satellites or the ground based station, the amount of training data (related to iterations $N$) fed to the GRU network can be increased or decreased. 

Fig. \ref{fig:time_history} shows the evolution of angular rates and Euler angles across four iterations (i.e, $N=4$). In the first iteration, both angular rates and Euler angles exhibit significant fluctuations over time. As the iterations progress, there is a notable reduction in the amplitude of these fluctuations, particularly from the second iteration onward. This is mainly because the GRU network is able to effectively capture the trend in the external disturbances acting on the satellites. Finally, by the fourth iteration, it is observed that the relative attitude and angular rate errors have reached a steady-state value. This trend verifies the efficacy of our proposed approach in mitigating the external disturbances acting on the GRACE satellites, resulting in a more precise stabilization of the relative attitude and angular rate errors over time compared to traditional PID controllers. The presence of a few spikes in the later iterations might be mainly due to attitude changes caused by the activation of the thrusters on these satellites (i.e, when the magnetic field $\mathbf{B}$ is parallel to the applied control input $\bu_{\text{PID}}$), but they have a minimal impact compared to the first iteration, highlighting the improved robustness and reliability of our proposed approach. Fig. \ref{fig:psd} shows the spectral density function of the relative attitude and angular rates at different iteration steps. This plot computes the power spectral density (PSD) via the fast fourier transform method. As seen from this figure, the maximum peak in these plots decreases with increase in the number of iterations. 

\begin{figure}[H]
    \centering
    \begin{subfigure}[b]{0.45\textwidth}
        \centering
        \includegraphics[width = 8.4 cm]{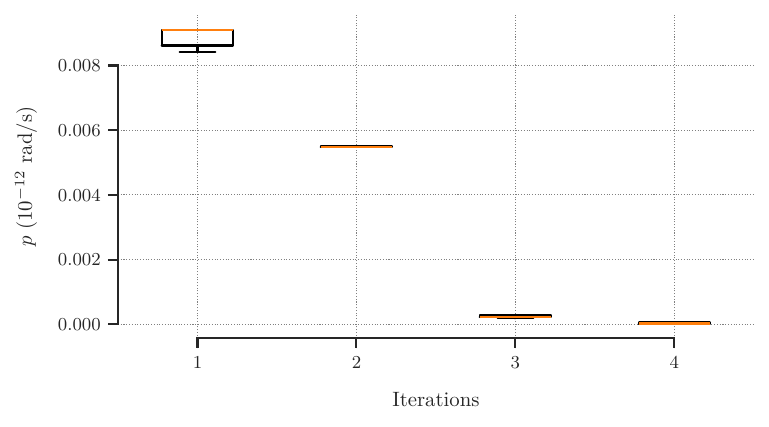}
    \end{subfigure}
    \hfill
    \begin{subfigure}[b]{0.45\textwidth}
        \centering
        \includegraphics[width = 8.4 cm]{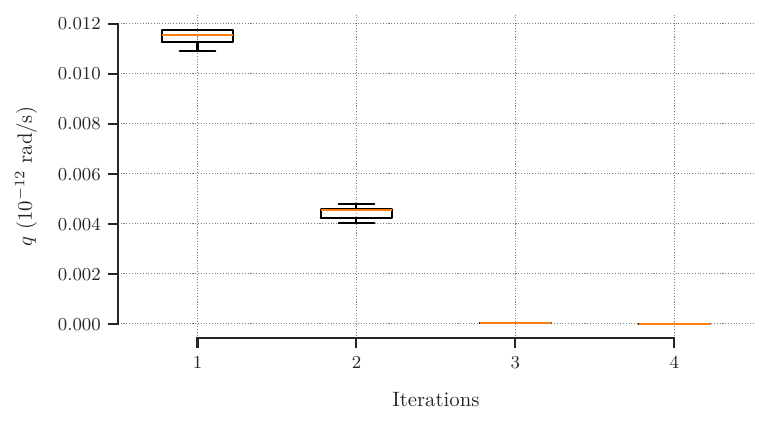}
    \end{subfigure}
    \begin{subfigure}[b]{0.45\textwidth}
        \centering
        \includegraphics[width = 8.4 cm]{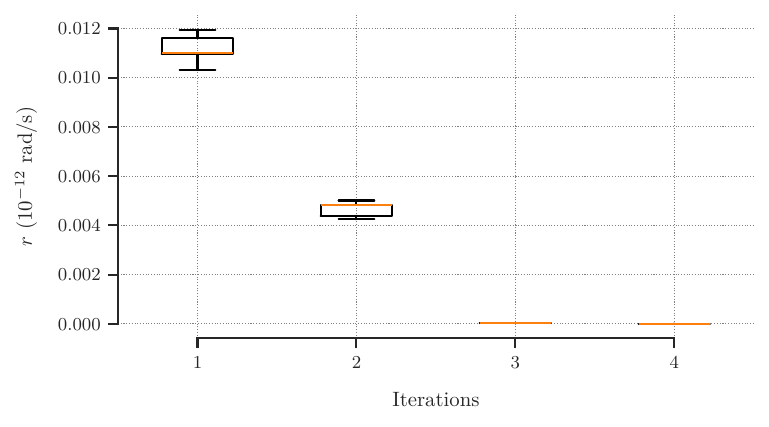}
    \end{subfigure}
    \hfill
    \begin{subfigure}[b]{0.45\textwidth}
        \centering
        \includegraphics[width = 8.4 cm]{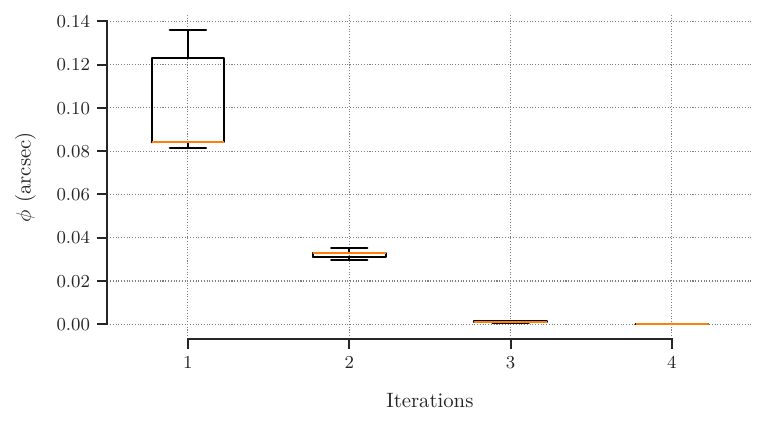}
    \end{subfigure}

    \begin{subfigure}[b]{0.45\textwidth}
        \centering
        \includegraphics[width = 8.4 cm]{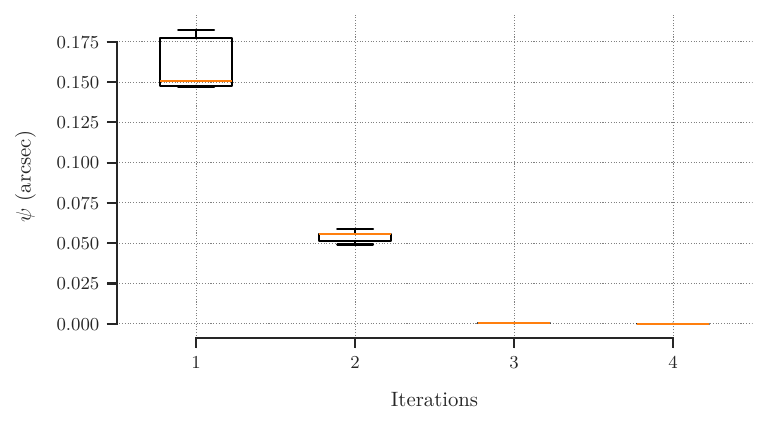}
    \end{subfigure}
    \hfill
    \begin{subfigure}[b]{0.45\textwidth}
        \centering
        \includegraphics[width = 8.4 cm]{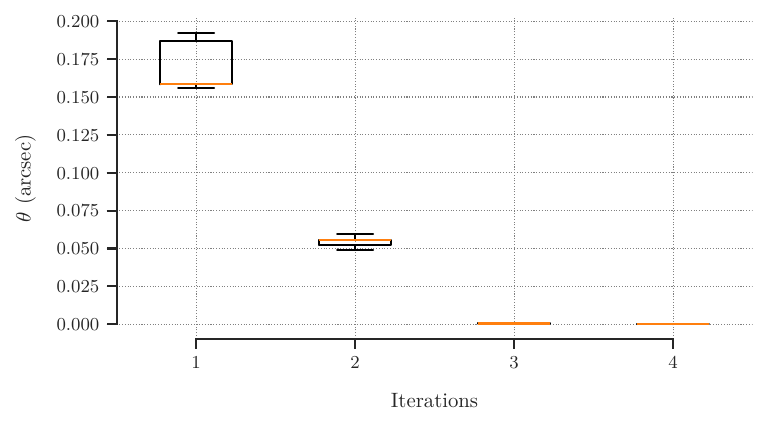}
    \end{subfigure}
    \caption{Box plots of relative attitude ($[\phi,\;\theta,\;\psi]$) and angular rate $\omega=[p,\; q,\; r]^\mathrm{T}$ error across four iterations for GRACE satellites}
    \label{fig:box_plot}
\end{figure}
\begin{figure}[H]

    \centering
    
    \begin{subfigure}[b]{0.45\textwidth}
        \centering
        \includegraphics[width=\linewidth]{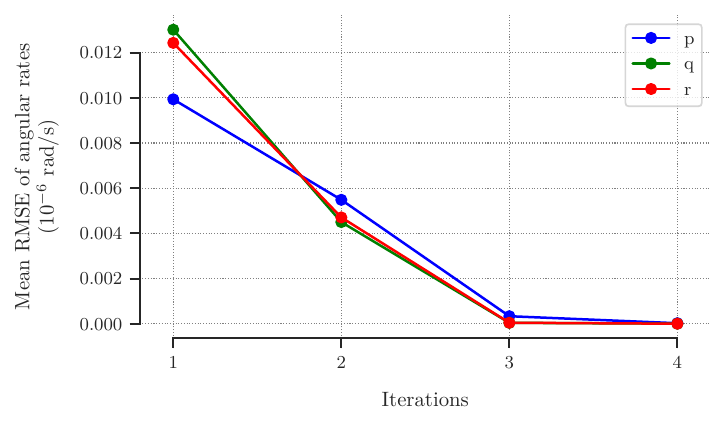}
        \label{fig:rmse_angular_rates}
    \end{subfigure}
    \begin{subfigure}[b]{0.45\textwidth}
        \centering
        \includegraphics[width=\linewidth]{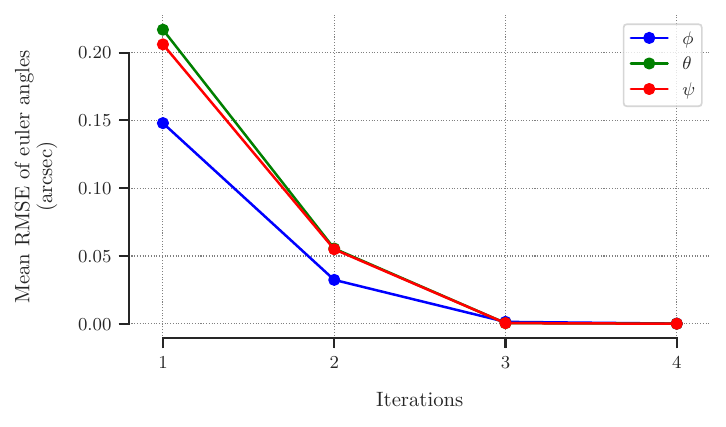}
        \label{fig:rmse_euler_angles}
    \end{subfigure}
    
    \caption{Evolution of mean RMSE with iterations}
    \label{fig:rmse_iteration}   
\end{figure}
Finally, the box plots in Fig. \ref{fig:box_plot} show the error distributions for the attitude angles ($\phi,\;\theta,\;\psi$) and angular rates ($p,\; q,\; r$) when the network has been trained multiple times over four iterations which are part of the proposed algorithm (Algorithm \ref{alg:proposed_approach}) for the GRACE-FO satellites. For attitude angles ($\phi,\;\theta,\;\psi$), the median values appear to stabilize or slightly decrease across iterations, with the interquartile range (IQR) narrowing, indicating a convergence towards a more precise estimate of the attitude angles. 
Fig. \ref{fig:rmse_iteration} depicts the evolution of the mean RMSE values for these variables with iterations. We can observe that these values continue to decrease with iterations and finally converge.

\section{Conclusion\label{sec:conclusion}}

In this paper, we explore the potential of Gated Recurrent Unit (GRU) networks for precise attitude and angular rate control, particularly for geodetic missions such as the GRACE-FO mission. Leveraging the relative attitude and angular rate time series data from the GRACE-FO data product, we estimated the additive time-varying disturbances acting on the satellites. Next, we observed a periodic trend in these disturbances data, prompting the utilization of GRUs to capture and predict these disturbance values. 
Our proposed modified PID controller minimally modifies the traditional PID controller by augmenting it with a GRU network that compensates for the disturbances acting on the satellites. Future work includes exploration of angular rate spectrum control and stability margins for the proposed controllers. Additionally, it would be interesting to investigate the susceptibility to sensor/actuator imperfections on the proposed controller.

\section{Acknowledgments}
The research has been supported by NASA Grant 80NSSC22K0287.

\bibliography{main}

\end{document}